\documentclass{article}
\usepackage{spconf,amsmath,graphicx}
\usepackage{booktabs}
\usepackage{caption}
\usepackage{hyperref}
\usepackage{amssymb}

\DeclareMathOperator*{\argmin}{arg\,min}
\usepackage{bm}
\usepackage{multicol, multirow}
\usepackage{comment}
\usepackage{enumitem}
\usepackage{tikz}
\usepackage{textcomp}
\usepackage{xpatch}

\def\L{{\cal L}}

\title{Weight Averaging: A Simple Yet Effective Method to Overcome Catastrophic Forgetting in Automatic Speech Recognition}
\name{Steven Vander Eeckt, Hugo Van hamme}
\address{KU Leuven \\ 
     Department Electrical Engineering ESAT-PSI, Leuven, Belgium\\
     \textit{\{steven.vandereeckt, hugo.vanhamme\}@esat.kuleuven.be}}
\newcommand\blfootnote[1]{%
  \begingroup
  \renewcommand\thefootnote{}\footnote{#1}%
  \addtocounter{footnote}{-1}%
  \endgroup
}

\begin{document}
\ninept
\maketitle
\begin{abstract}
\blfootnote{Research supported by Research Foundation Flanders (FWO) under grant S004923N of the SBO programme.}
Adapting a trained Automatic Speech Recognition (ASR) model to new tasks results in catastrophic forgetting of old tasks, limiting the model's ability to learn continually and to be extended to new speakers, dialects, languages, etc. Focusing on End-to-End ASR, in this paper, we propose a simple yet effective method to overcome catastrophic forgetting: weight averaging. By simply taking the average of the previous and the adapted model, our method achieves high performance on both the old and new tasks. It can be further improved by introducing a knowledge distillation loss during the adaptation. We illustrate the effectiveness of our method on both monolingual and multilingual ASR. In both cases, our method strongly outperforms all baselines, even in its simplest form. 
\end{abstract}
\begin{keywords}
end-to-end automatic speech recognition, continual learning, conformer, transformer, weight averaging
\end{keywords}

\section{Introduction}
\label{sec:intro}
Catastrophic Forgetting (CF) \cite{catastrophicforgetting} remains a problem when adapting a trained Automatic Speech Recognition (ASR) model to new tasks, regardless of whether the new tasks are new languages, dialects, accents or simply new speakers. CF thus severely limits the ability of ASR models to be extended to new domains (e.g. to build very powerful ASR models able to perform well for all dialects, accents, speakers, etc.), or to exploit all the (recently become available) data at its disposal. To overcome CF, one needs to re-introduce all past data when extending the ASR models, which soon becomes very expensive in terms time, energy and resources.

Alternatively, one could use Continual Learning (CL) methods to enable models to learn continually without suffering from CF. CL has been a hot research topic in image classification, with many CL methods being proposed. These CL methods can be grouped into three categories \cite{defy}: (i) regularization-based methods add a term to the loss to regularize training, e.g. \cite{ewc, mas, imm}; (ii) rehearsal-based methods rehearse previous tasks through a small memory of samples from previous tasks, e.g. \cite{agem, er}; (iii) architectural-based methods increase the model capacity when learning new tasks, e.g. \cite{pnn}. For ASR, CL is still a relatively new topic. \cite{lifelongasr} and \cite{eeckt2021continual} implement, respectively, four and nine CL methods for (End-to-End) E2E ASR, both finding that rehearsal-based methods remain the most practical way to overcome CF.  \cite{ogem} applies Gradient Episodic Memory \cite{gem}, also a rehearsal-based method, to E2E ASR, while assuming that task boundaries are not known.  However, the disadvantage of rehearsal-based methods is that they require storing a memory (i.e. a set of samples from previous tasks), which may not always be allowed due to privacy concerns. \cite{eeckt_adapters} finds that using rehearsal-based methods is not necessary to prevent CF when task-specific adapters \cite{adapters} are used, which comes, however, at the cost of introducing task-specific parameters, possibly requiring a task label at inference time. 

Similar to \cite{lifelongasr, eeckt2021continual, eeckt_adapters}, this paper focuses on CL (with known task boundaries) for E2E ASR, and proposes a simple yet very effective method to overcome CF: weight averaging. By simply computing the (weighted) average of the model before and the model after adaptation to the new task, CF can be prevented while learning the new task well. Though already very effective by itself, weight averaging can be further improved by introducing knowledge distillation losses (as in \cite{lwf}) during the adaptation. Overall, our method remains very simple, while, compared to the best methods from \cite{lifelongasr, eeckt2021continual}, not requiring a memory.  We illustrate the effectiveness of our method on both monolingual and multilingual experiments, consisting of six and five tasks, respectively. On both experiments, our method strongly outperforms all baselines, even in its simplest form.

\section{ASR Model}
Our ASR model is an E2E encoder-decoder model consisting of a Conformer encoder \cite{conformers} and a Transformer decoder \cite{transformer}. The output of the model are $o$ word pieces, and its loss is a combination of the decoder cross-entropy (CE) loss and a CTC loss computed after applying a linear layer to the encoder output. 
We refer to the model parameters as $\theta \in \mathbb{R}^N$ with $N$ the number of parameters.  $f^\text{ctc}(X; \theta)$ and $f^\text{dec}(X; \theta)$ are the outputs (after applying the softmax) of the CTC and decoder, respectively, for an input utterance $X \in \mathbb{R}^{L \times d_i}$, which consists of $L$ input frames. Given $\L^\text{ctc}(X, y; \theta)$ the CTC loss and $\L^\text{dec}(X, y; \theta)$ the decoder CE loss for utterance $X$  with $y$ its ground truth (of $W$ word pieces), the loss of the model is given by:
\begin{equation}
    \L_\text{ce}(X, y; \theta) = \alpha\L_\text{ctc}(X, y; \theta) + (1 - \alpha)\L_\text{dec}(X, y; \theta)
    \label{eq:loss}
\end{equation}

\section{Continual Learning}
\label{sec:cl}
There are $T$ tasks which must be learned in sequence (e.g. because they become available at different times). For each task $t$, we have access to its train set $D^\text{train}_t$ and validation set $D^\text{val}_t$, which both contain pairs $(X, y)$. Importantly, the access to $D^\text{train}_{t-1}$ and $D^\text{val}_{t-1}$ of previous task $t-1$ is assumed to be lost once it has been learned by the model and the new task $t$ becomes available. 
Consequently, given a model trained on task $t-1$ with parameters $\theta^{t-1}$, the objective of Continual Learning is to be able to train this model on a new task $t$, obtaining parameters $\theta^{t}$ while satisfying two criteria:  
\begin{enumerate}[label=(\alph*)]
    \item retain high performance on all previous tasks $1$ to $t-1$. Ideally, thus, for each task $j \in \{1, ..., t-1\}$:
    \begin{equation}
        \sum_{(X, y) \in D^\text{train}_j}  \L_\text{ce}\left(X, y; \theta^{t}\right) \leq \sum_{(X, y) \in D^\text{train}_j} \L_\text{ce}\left(X, y; \theta^{t-1}\right)
        \label{eq:knowledge_retention}
    \end{equation}
    This criterion is referred to as \textit{knowledge retention} \cite{biesialska-etal-2020-continual}. The difficulty lies in the fact that $D_j^\text{train}$ is no longer available.
    \item achieve high performance on the new task $t$. Thus, $\theta^{t}$ should satisfy (within early stopping):
    \begin{equation}
        \theta^{t} = \argmin_\theta \sum_{(X, y) \in D^\text{train}_{t}}  \L_\text{ce}\left(X, y; \theta \right)
        \label{eq:forward_transfer}
    \end{equation}
    This is referred to as \textit{forward transfer} \cite{biesialska-etal-2020-continual}.
\end{enumerate}
Note that in case the parameters $\theta$ of the model can be written as $\theta=\{\theta^S, \theta^U\}$ with $\theta^S$ the parameters shared across tasks and $\theta^{U}$ the task-specific parameters, $\theta^{t}=\{\theta^{S,t}, \theta^{U,1},...,\theta^{U,t}\}$ with $\theta^{U,i}$ the task-specific parameters for task $i \leq t$. To transcribe an utterance $X$ of task $i \leq t$, parameters $\{\theta^{S,t}, \theta^{U,i}\}$ are then used.

Combined, the two criteria above require that after training on task $t$, the model must have high performance on all seen tasks $1$ to $t$. To achieve this, it has access to the parameters of a single model performing well on tasks $1$ to $t-1$, and $D^\text{train}_{t}$ and $D^\text{val}_{t}$ of the new task $t$. Since naively (i.e. fine-tuning) training the model with parameters $\theta^{t-1}$ on $D^\text{train}_{t}$ will result in catastrophic forgetting, we discuss in the remainder of this Section how our model combats CF. 

\subsection{Weight Averaging}
\label{subsec:weight_avg}
The major aspect of our approach is weight averaging \cite{Utans96weightaveraging}. Before training on task $t$, we have a model with parameters $\theta^{t-1}$. Next, we proceed in two steps. First, we train this model on task $t$ to obtain parameters $\hat{\theta}^{t}$, either by naively fine-tuning (which will result in catastrophic forgetting) or by using a CL method (see Sec. \ref{subsec:cl_method}). Second, we combine the parameters of the 'old' model ($\theta^{t-1}$) and those of the new model ($\hat{\theta}^{t}$) as follows:
\begin{equation}
    \theta^{t} = (1 - \eta) \theta^{t-1} + \eta \hat{\theta}^{t}, \: \: \: 0 \leq \eta \leq 1
    \label{eq:weight_avg}
\end{equation}
with $\theta^{t}$ the final parameters after learning task $t$. If $\theta=\{\theta^S, \theta^U\}$, i.e. there is a shared and task-specific part, Eq. \ref{eq:weight_avg} is applied to the shared part $\theta^S$ of $\theta^{t-1}$ and $\hat{\theta}^{t}$ only. Regarding $\eta$, we consider: (a) $\eta=0.5$; (b) $\eta=t^{-1}$. For (a), Eq. \ref{eq:weight_avg} can be written as an exponential moving average in which the contribution of $\tilde{\theta}^i$ goes exponentially to zero as $|i-t|$ increases. For (b), we can write Eq. \ref{eq:weight_avg} as $\theta^{t}=t^{-1}\sum_{i=1}^{t} \tilde{\theta}^i$, and this is thus an average over $\tilde{\theta}^i$ for all $i \leq t$. 

\subsection{Adding Knowledge Distillation Loss}
\label{subsec:cl_method}
As explained in Sec. \ref{subsec:weight_avg}, before averaging the weights, we first adapt the model with parameters $\theta^{t-1}$ to the new task $t$, obtaining a model with parameters $\hat{\theta}^{t}$. This adaptation can be done either naively (i.e. fine-tuning) or by using a CL method. For the latter case, we consider Learning Without Forgetting (LWF) \cite{lwf}, a regularization-based CL method. We consider LWF since it is a relatively simple method, which does not require a memory (i.e. it is not a rehearsal-based method) and is very storage efficient \cite{eeckt2021continual}. As shown in \cite{eeckt2021continual}, by itself, LWF is not strong enough to prevent CF in E2E ASR (though still works better than fine-tuning), but it might improve the CL performance of weight averaging while still keeping the overall method relatively simple and memory-free. 

When training task $t$ starting from the model with parameters $\theta^{t-1}$, LWF uses a knowledge distillation loss \cite{knowledge_distillation} between the 'old model' (with parameters $\theta^{t-1}$) and the current model:
\begin{equation}
    \begin{aligned}
    \L_\text{lwf} & (X; \theta) = \alpha \sum_{i=1}^L \sum_{j=1}^o {f_{i,j}^{ctc}(X; \theta^{t-1}_\text{old})} \log {f_{i,j}^{ctc}(X; \theta)} \\
     & + (1 - \alpha) \sum_{i=1}^W \sum_{j=1}^o {f_{i,j}^{dec}(X; \theta^{t-1}_\text{old})} \log {f_{i,j}^{dec}(X;\theta)} 
    \end{aligned}
    \label{eq:loss_lwf}
\end{equation}
with $f_{i,j}^{dec}$ the decoder probability of the $j$th word piece for the $i$th element in the output sequence of utterance $X$; $f_{i,j}^{ctc}$ is the same for CTC; and $\theta^{t-1}_\text{old}=\theta^{t-1}$ if the entire model is shared, otherwise $\theta^{t-1}_\text{old}=\{\theta^{S,t-1},\theta^{U,t}\}$, i.e. $X$, which belongs to task $t$, is sent through the old model's shared part but through the task-specific part of task $t$. LWF thus transfers knowledge from the old to the new model, and to this end it uses the new task's data. During training, the above loss is added to the loss from Eq. \ref{eq:loss}, so that the loss when training the model with parameters $\theta^{t-1}$ on task $t$ becomes:
\begin{equation}
    \L(X, y; \theta) = \L_\text{ce}(X, y; \theta) + \lambda \L_\text{lwf}(X; \theta)
    \label{eq:loss_cl}
\end{equation}
With $\lambda \geq 0$. After training with the above loss, we obtain parameters $\hat{\theta}^{t}$, which are then averaged with $\theta^{t-1}$, as explained in Sec. \ref{subsec:weight_avg}, to obtain $\theta^{t}$. We refer to our method as FTA (Fine-Tuning with Averaging) if $\lambda=0$ and LWFA (LWF with Averaging) if $\lambda > 0$.

\section{Experiments}
We consider two experiments (done in ESPnet \cite{watanabe2018espnet}): one monolingual, where the entire model is shared across tasks; and one multilingual, where the encoder is shared while the decoder and the linear layer of the CTC loss are made task-specific (i.e. language-specific). 

\begin{table*}
    \centering
    \caption{Results on the CV English experiments after learning the six tasks (i.e. dialects) in sequence. All WERs are expressed in percentages. WERs are reported per task, evaluated on the final model, in addition to the summary results (AVG, BWT and FWT). $\eta$ is the weight in weight averaging (see Sec. \ref{subsec:weight_avg}). 'Mem.' indicates the number of utterances stored in a memory, if applicable. Best model is highlighted in bold.}
    \begin{tabular}{l l c c c c c c c r r r}
    & & & \multicolumn{6}{c}{\textit{WER per task}} & \multicolumn{3}{c}{\textit{Summary}} \\
    \cmidrule(lr){4-9} \cmidrule(lr){10-12} 
    Model & \multicolumn{1}{c}{$\eta$} & Mem. & T1--US & T2--ENG & T3--AUS & T4--IND & T5--SCO & T6--IRE & \textbf{AVG} & \textbf{BWT} & \textbf{FWT}  \\
    \toprule
    Sep. Model & & & 17.3 & 10.8 & 10.6 & 16.7 & 12.1 & 11.4 & 13.14 & 0.0 & 0.0 \\
    Fine-Tuning & & & 19.4 & 12.7 & 14.0 & 20.6 & 13.4 & 11.4 & 15.25 & -2.5 & 0.0  \\
    \midrule 
    KD & & 0.5k & 18.4 & 12.2 & 13.7 & 20.2 & 12.8 & 10.8 & 14.67 & -2.0 & +0.2  \\
    ER & & 0.5k & 18.4 & 12.1 & 13.6 & 18.8 & 13.0 & 11.1 & 14.49 & -1.6 & +0.0  \\
    ER & & 2.0k & 18.1 & 12.0 & 13.0 & 18.6 & 12.7 & 10.8 & 14.20 & -1.3 & +0.1  \\
    LWF & & & 19.0 & 12.2 & 13.7 & 20.7 & 12.6 & 11.1 & 14.88 & -2.3 & +0.2  \\
    \midrule
    FTA & $=0.50$ & & 17.5 & 11.4 & 12.8 & 19.6 & 12.0 & 10.4 & 13.94 & -0.8 & -0.1 \\
     & $=t^{-1}$ & & 17.2 & 11.1 & 12.0 & 19.3 & 12.4 & 10.4 & \textbf{13.72} & -0.1 & -0.5   \\   
    LWFA & $=0.50$ & & 17.2 & 11.2 & 12.7 & 19.2 & 11.9 & 10.5 & 13.79 & -0.7 & +0.0  \\
     & $=t^{-1}$ & & 17.1 & 11.1 & 11.9 & 19.5 & 12.4 & 10.3 & \textbf{13.72} & -0.1 & -0.5 \\
    \bottomrule
    \end{tabular}
    \label{tab:mono}
\end{table*}

\textbf{Data.} For the experiments, we consider the Common Voice (CV) \cite{commonvoice} dataset (version 7.0). For the monolingual experiments, we consider CV English, which we split into 6 dialects: United States (US), England (ENG), Australia (AUS), India (IND), Scotland (SCO), Ireland (IRE). The tasks (i.e. dialects) are learned in this order, which was randomized, except that we start with the largest task, followed by the second largest task, which seems the most realistic scenario in practice. The number of utterances ranges from approximately 350k for US to 7k for IRE. For the multilingual experiments, the first task is the same as for the monolingual experiments, i.e. US; then, we learn in the following order: Dutch (NL), Swedish (SV), Polish (PL) and Russian (RU). The ordering is based on the languages' similarity to English (US), with similar languages coming first. In this case, the smallest task is SV with 23k utterances. 

\textbf{Model.} The model contains 12 Conformer encoders and 6 Transformer decoders, with attention dimension $256$ and feedforward dimension 2048. The output dimension $o$ equals 5000 word pieces generated with \cite{sentencepiece}. The number of parameters equals 46.8M. The optimizer is Adam \cite{adam}, reset after each task. The learning rate is ten times smaller for subsequent tasks than for the initial one. The final model after learning a task is obtained by averaging ten best checkpoints \cite{transformer}.   For the monolingual experiments, the entire model is shared, including the output layers and the word pieces generated on the first task (US). For the multilingual experiments, the decoder, the output layers and the word pieces are language-specific, which means that only the encoder and thus $71.5\%$ of the model is shared. Moreover, for all multilingual models, learning a new language is done in two steps: 1) we freeze the encoder and only train the task-specific layers; 2) we train the entire model. 

\textbf{Baselines.} We consider the following baselines: (i) Fine-Tuning: naively adapts the model to new tasks, suffering from CF; (ii): Separate Model (Sep. Model): same as Fine-Tuning, but stores all past models $\theta^{i}$ for task $i$, so that each task has its separate model (i.e. after $t$ tasks it stores $t$ models) -- this violates the idea of CL but can be considered an upper bound; (iii) Experience Replay (ER) \cite{er}: rehearsal-based method, which trains jointly on a mini-batch from the new task and one sampled from the memory set; (iv) Knowledge Distillation (KD): rehearsal-based method as used in \cite{eeckt2021continual}, in which it was the best CL method - it uses the loss of LWF (Eq. \ref{eq:loss_lwf}), but computed on a mini-batch from the memory set; (v) LWF \cite{lwf}: best regularization-based method from \cite{eeckt2021continual}. For the multilingual experiments, we add Freeze Encoder (Freeze Enc.), which freezes the encoder after training on US. For the rehearsal-based methods, a memory with fixed size $M$ (i.e. $M/t$ utterances per task for $t$ seen tasks) is formed by sampling uniformly from each task's training set. We consider $M=0.5k$ and $M=2.0k$. For all methods, including LWF in LWFA, the hyper-parameters from \cite{eeckt2021continual} are used. 

\textbf{Metrics.} We use the same metrics as \cite{eeckt2021continual}, based on WER: Average WER (AVG) is the average WER over all seen tasks, evaluated on the final model; Backward Transfer (BWT) \cite{gem} is the average decrease in WER of previous tasks since they were first learned - negative BWT is forgetting; Forward Transfer (FWT) \cite{gem} is the average decrease in WER on each new task compared to Fine-Tuning, so that a positive FWT means better learning than Fine-Tuning. It is clear that AVG is the main metric, and whether one method outperforms another is determined by their AVG. However, BWT and FWT can give an idea to which extent the model satisfies the criteria \textit{knowledge retention} and \textit{forward transfer} from Sec. \ref{sec:cl}.

\section{Results}

\begin{table*}
    \centering
    \caption{Results on the CV Multilingual experiments after learning the five tasks (i.e. languages) in sequence. All WERs are expressed in percentages. WERs are reported per task, evaluated on the final model, in addition to the summary results (AVG, BWT and FWT). $\eta$ is the weight in weight averaging (see Sec. \ref{subsec:weight_avg}). 'Mem.' indicates the number of utterances stored in a memory. Best model is highlighted in bold.}
    \begin{tabular}{l l c c c c c c r r r }
    & & & \multicolumn{5}{c}{\textit{WER per task}} & \multicolumn{3}{c}{\textit{Summary}} \\
    \cmidrule(lr){4-8} \cmidrule(lr){9-11} 
    Model & \multicolumn{1}{c}{$\eta$} & Mem. & T1--US & T2--NL & T3--SV & T4--PL & T5--RU & \textbf{AVG} & \textbf{BWT} & \textbf{FWT} \\
    \toprule
    Sep. Model &  & & 17.3 & 10.5 & 38.0 & 8.1 & 11.0 & 16.97 & 0.0 & 0.0  \\
    Freeze Enc. & & & 17.3 & 17.9 & 55.1 & 13.8 & 21.2 & 25.06 & 0.0 & -10.1  \\
    Fine-Tuning & & & 59.4 & 30.7 & 61.7 & 10.9 & 11.0 & 34.73 & -22.2 & 0.0  \\
    \midrule 
    KD & & 0.5k & 37.5 & 36.1 & 59.8 & 13.1 & 12.2 & 31.71 & -18.1 & -0.3  \\
    ER & & 0.5k & 36.8 & 20.4 & 57.4 & 11.2 & 12.4 & 27.64 & -12.9 & -0.4\\
    ER & & 2.0k & 31.0 & 16.0 & 49.3 & 9.9 & 12.0 & 23.65 & -8.1 & -0.3 \\
    LWF  & & & 54.4 & 34.3 & 57.2 & 8.9 & 10.0 & 32.93 & -21.1 & +1.2  \\
    \midrule
    FTA & $=0.50$ & & 28.1 & 15.6 & 44.2 & 8.4 & 12.5 & 21.76 & -4.8 & -1.2 \\
    & $=t^{-1}$ & & 21.2 & 13.3 & 41.2 & 10.2 & 15.5 & 20.28 & -1.2 & -2.9  \\
    LWFA & $=0.50$ & & 26.8 & 14.8 & 43.7 & 8.2 & 12.1 & 21.12 & -4.4 & -0.8 \\
     & $=t^{-1}$ & & 20.4 & 12.7 & 39.6 & 10.1 & 14.8 & \textbf{19.54} & -0.9 & -2.3  \\
    \bottomrule
    \end{tabular}
    \label{tab:multi}
\end{table*}

\subsection{Monolingual Experiments}

Table \ref{tab:mono} shows the results of the monolingual experiments, where the methods must learn six dialects of English in sequence. 

First, note the effectiveness of weight averaging for the monolingual tasks. All settings of our method outperform all baselines, including ER with 2.0k utterances, while our methods do not even require storing a memory. Note also how Fine-Tuning and LWF, which are the same as FTA and LWFA without weight averaging, suffer from forgetting and are unable to perform well.

Second, considering our methods, we find, on the one hand, that for $\eta$ (see Sec. \ref{subsec:weight_avg}), $\eta=t^{-1}$ works better than $\eta=0.50$. In the latter, the contribution of an adapted model $\hat{\theta}^i$ for task $i$ in the current model $\theta^{t+1}$ goes exponentially to zero as $|i-t|$ increases. Consequently, it should not be a surprise that $\eta=0.50$, though it learns the new task very well (FWT), suffers from slight forgetting. Nevertheless, given that after six tasks, the contribution of the model trained on T1--US is less than $2\%$ (i.e. $0.50^{6}$), it is surprising to see that even for $\eta=0.50$, our methods suffer from less forgetting than the baselines and retain higher WER on the first three tasks.
Thus, even for relatively high $\eta$, weight averaging is able to alleviate forgetting. When $\eta=t^{-1}$, each adapted model $\hat{\theta}^i$ for task $i$ contributes equally in the current model $\theta^{t+1}$. This results in slightly poorer learning of new tasks (FWT), but also almost zero forgetting (BWT), for both FTA and LWFA. Overall, for both FTA and LWFA, $\eta=t^{-1}$ is a better option than $\eta=0.50$. On the other hand, comparing LWFA to FT, we find that LWFA improves the performance of FTA only for $\eta=0.50$. Since for $\eta=0.50$, the contribution of older models goes exponentially to zero, it is not surprising to find that LWFA, which contains a mechanism to reduce forgetting during the adaptation, retains more old knowledge than FTA. Indeed, for $\eta=0.50$, LWFA outperforms FTA for all of the five previous tasks. However, the need for a CL method during the adaptation is overcome when using $\eta=t^{-1}$. In that case, an equal focus is put on all tasks and the mechanism in LWFA that reduces forgetting during the adaptation does not improve the performance.

Overall, we can conclude that FTA and LWFA with $\eta=t^{-1}$ (and to a slightly lesser extent with $\eta=0.50$) are extremely effective and enable CL in monolingual ASR, approximately achieving \textit{knowledge retention} and \textit{forward transfer} (Sec. \ref{sec:cl}). To achieve their high performance, they do not even require a memory, unlike the strongest baselines. They improve the strongest baseline, ER with 2.0k utterances, by 3.4$\%$, without requiring a memory. Compared to baselines storing only 0.5k utterances or even zero utterances, our method improves their performance by 5.3$\%$ and 7.8$\%$, respectively.

\subsection{Multilingual Experiments}
However, it might be that weight averaging only works for similar tasks, which is the case for the monolingual experiments. To this end, we test our method on the multilingual experiments, in which the tasks are much more dissimilar. Table \ref{tab:multi} shows the results. 

First, it is indeed clear that forgetting is an even more severe problem for the multilingual experiments, where the tasks are more dissimilar, than for the monolingual experiments. In particular for Fine-Tuning and LWF, the forgetting can indeed be considered catastrophic. However, as for the monolingual experiments, our method still proves to be extremely effective in reducing CF, strongly outperforming all baselines, including ER with 2.0k utterances. The latter, using a large memory, and our methods are the only methods outperforming Freeze Enc., which freezes the encoder after US.

Second, considering our methods, for the value of $\eta$ (Sec. \ref{eq:weight_avg}), the same conclusions can be drawn as for the monolingual experiments: $\eta=t^{-1}$ is better than $\eta=0.50$, with much less forgetting (BWT) and only slightly worse FWT; nevertheless, $\eta=0.50$ performs surprisingly well on old tasks (better than the baselines). Regarding LWFA compared to FTA, we find that, while in the monolingual experiments the advantage of LWFA over FTA was very small, in the multilingual experiments, where the forgetting is more severe, the advantage of using LWFA over FT is much more significant. 

Overall, FTA and LWFA with $\eta=t^{-1}$ reduce the forgetting of ER with 2.0k utterances by, respectively, 85.2$\%$ and $88.9\%$, while not requiring a memory. Given that their FWT is only slightly worse than the latter's, they improve the latter's performance by 14.2$\%$ and 17.4$\%$, respectively. If the number of utterances that can be stored is further limited to 0.5k or even zero, LWFA improves the strongest baseline by, respectively, 29.3$\%$ and 40.7$\%$. In addition, it improves Freeze Enc., overall the second strongest method, by 22.0$\%$. Consequently, FTA and especially LWFA can again be said to have approximately achieved \textit{knowledge retention} and \textit{forward transfer}. 

\section{Discussion}
In checkpoint averaging, the final model is obtained by averaging over the most recent or best performing checkpoint. Checkpoint averaging is commonly associated with Transformers \cite{transformer}. \cite{swa} argues that averaging all the checkpoints along the trajectory of stochastic gradient descent leads to wider optima, and, consequently, better generalization. Checkpoint averaging has been used by \cite{stahlberg-etal-2019-cued}  which combines averaging over checkpoints stored every 500 iterations with Elastic Weight Consolidation \cite{ewc}, to overcome overfitting when fine-tuning a pre-trained model to a target domain. \cite{fast_checkpoint} proposes to average the most recent checkpoints after each epoch during training, in order to reduce training time. 

More broadly, averaging weights of Artificial Neural Networks (ANN) has been used as early as 1996, by \cite{Utans96weightaveraging}, in which it is shown that averaging over ANNs trained on the same task but different data, and from the same pre-trained model, reduces the variance. Similar work, focusing on neural machine translation, is \cite{ensemble_averaging}, which computes the average weights of four models in an ensemble, finding that the resulting model nearly matches the performance of the ensemble itself. \cite{model_soups} introduces 'model soups', which average the weights of multiple models trained on the same task and from the same initialization but with different hyper-parameter configurations. \cite{robust_fine_tuning} uses weight averaging between a pre-trained zero-shot model and the same model fine-tuned on distribution shifts to improve the robustness of the model and achieve high performance on the target domain as well as on the distribution shifts. For an explanation of why this works, they consider  \cite{NEURIPS2020_0607f4c7}, which observes that "pre-trained weights guide the optimization to a flat basin of the loss landscape." 

Our experimental observations are compatible with the explanation from \cite{swa, NEURIPS2020_0607f4c7} that weight averaging leads to wider optima and hence better generalization.

\section{Conclusion}

In this paper, we proposed weight averaging as a simple yet effective method to overcome catastrophic forgetting in E2E ASR. Our method proceeds in two steps: 1) it trains the model on the new task, either by fine-tuning or using LWF; 2) it computes the average of the old model and the new model with weights $1-\eta$ and $\eta$, respectively. We illustrated the effectiveness of weight averaging on both monolingual and multilingual experiments, on which our best method outperformed the best baseline by $3.4\%$ and $17.4\%$, respectively. This baseline required storing a memory of 2.0k utterances, while our methods do not. Consequently, our method is able to achieve very high CL performance, without even requiring a memory, which was previously found to be necessary for CL in E2E ASR to work well. Regarding $\eta$, we found that $\eta=t^{-1}$ for after $t$ tasks works best, resulting in nearly zero forgetting on old tasks while learning new task well. Using LWF instead of fine-tuning during the first step of our method improved the performance only in the multilingual experiments, in which the tasks are more dissimilar; in the monolingual experiments, fine-tuning worked equally well. 
 
\bibliographystyle{IEEEbib}
\bibliography{main}

\end{document}